\documentclass[11pt,a4paper]{article}
\usepackage[margin=1in]{geometry}
\usepackage[T1]{fontenc}
\usepackage[utf8]{inputenc}
\usepackage{lmodern}
\usepackage{amsmath,amssymb}
\usepackage[numbers,sort&compress]{natbib}
\usepackage{graphicx}
\usepackage{array,booktabs,multirow}
\usepackage{microtype}
\usepackage{authblk}
\usepackage[hidelinks]{hyperref}

\usepackage[labelfont=bf,labelsep=period]{caption}
\title{PaECTER: Patent-level representation learning using citation-informed transformers}
\author[1]{Mainak Ghosh}
\author[1]{Michael E. Rose\thanks{Corresponding author: \href{mailto:michael.rose@ip.mpg.de}{michael.rose@ip.mpg.de}}}
\author[1]{Sebastian Erhardt}
\author[1]{Erik Buunk}
\author[1,2,3]{Dietmar Harhoff}
\affil[1]{Innovation \& Entrepreneurship Research, Max Planck Institute for Innovation and Competition, Munich, Germany}
\affil[2]{Ludwig Maximilian University of Munich, Munich, Germany}
\affil[3]{Centre for Economic Policy Research (CEPR), London, United Kingdom}
\date{\small Published in \textit{PLOS ONE} 21(10): e0358188 (2026), \href{https://doi.org/10.1371/journal.pone.0358188}{doi:10.1371/journal.pone.0358188}}

\begin{document}
\maketitle

\begin{abstract}
PaECTER is an open-source document-level encoder specific to patents.
We fine-tune BERT for Patents with examiner-added citation information to generate numerical representations for patent documents.
PaECTER performs better in similarity tasks than current state-of-the-art models used in the patent domain.
More specifically, our model outperforms the patent specific pre-trained language model (BERT for Patents) and general-purpose text embedding models (e.g., E5, GTE, and BGE) on our patent citation prediction test dataset on different rank evaluation metrics.
PaECTER predicts at least one most similar patent at a rank of 1.32 on average when compared against 25 irrelevant patents.
Numerical representations generated by PaECTER from patent text can be used for downstream tasks such as classification, tracing knowledge flows, or semantic similarity search.
Semantic similarity search is especially relevant in the context of prior art search for both inventors and patent examiners.
\end{abstract}

\section*{Introduction}

Domain-specific language models have become increasingly critical in the social sciences and technological domains, enabling the transformation of text into valuable data and offering scalable ways to learn semantic similarity across large document collections. In the context of patents, social scientists have shown particular interest in measuring patent similarity to analyze knowledge flows, estimate patent value, and study technological frontiers. For patent examiners, the ability to retrieve relevant prior art precisely and efficiently is essential for timely assessments of the patentability of claimed inventions under patent law and ensuring the quality of patent examinations. Missing any relevant prior patents during prior art searches would risk granting suboptimal patents, which can result in opposition or litigation, increase the burden of rework for the patent office, and undermine the objectives of the patent system.

However, existing state-of-the-art tools for document similarity learning are often specifically tailored to scientific English, which structurally and lexically differs from patent language. The PaECTER model (Patent Embeddings using Citation-informed TransformERs) addresses this gap by encoding patents into numerical representations to enable the identification of semantically similar patents.

The introduction of Bi-Directional Encoder Representations from Transformers (BERT)~\cite{devlin_bert_2019} marked a significant advancement for researchers working with text. These models, trained on large textual corpora, produce fixed-size dense vector representations (embeddings) of input text. BERT and its derivatives rely on a predefined vocabulary of tokens, which are either words or sub-words. Out-of-vocabulary tokens are either discarded or broken into smaller, recognizable subwords. This approach can result in incomplete linguistic coverage, particularly in domains with specialized language, motivating the development of domain-specific BERT derivatives.

Researchers have responded by developing BERT models tailored to specific domains and corpora. Among them, BERT for Patents~\cite{srebrovic_leveraging_2020} and SciBERT~\cite{beltagy_scibert_2019} are pre-trained language models optimized for patent texts and scientific language, respectively. Vocabulary selection plays a crucial role in determining the effectiveness of these models for domain-specific tasks. Our analysis highlights this by showing that only approximately 50\% of the vocabulary used in SciBERT overlaps with that of BERT for Patents --- underscoring the need for specialized models in patent-related research.

However, these models face a limitation when it comes to measuring text similarity. Their training objectives --- predicting masked tokens and the next sentence --- are not inherently designed for identifying similar documents.

To address this limitation, it is necessary to define an appropriate similarity signal and adopt a similarity-focused training objective. Contrastive learning is a machine learning approach well-suited for this purpose. It trains models to identify patterns by emphasizing the recognition of similarities and differences between data points. A key component of this method is the triplet loss function, which relies on specially crafted training data containing explicit signals for similarity and dissimilarity.

SPECTER~\cite{cohan_specter_2020} and SPECTER2~\cite{singh_scirepeval_2023} exemplify the application of this approach. Both models build on SciBERT’s ability to represent scientific language and utilize training datasets composed of triplets. Each triplet includes a query, a related document (e.g., a citation), and an unrelated one (e.g., a non-citation). The original SPECTER model was trained on 684,000 triplets derived from 146,000 query documents, while the more advanced SPECTER2 was trained on over 6 million triplets. By encoding the titles and abstracts of scientific publications, both models have demonstrated superior performance on various downstream tasks in the scientific domain, including topic classification, citation prediction, and recommendation.

SEARCHFORMER~\cite{vowinckel_searchformer_2023} applies a similar approach in the patent domain. However, it is designed specifically for comparing patent claims, making it unsuitable for tasks focused on patent titles and abstracts. Additionally, the model is not publicly available.

\begin{figure}
  \centering
  \caption{Relationship of PaECTER model to existing models}\label{fig:bert-and-specters}
  \includegraphics[width=7cm]{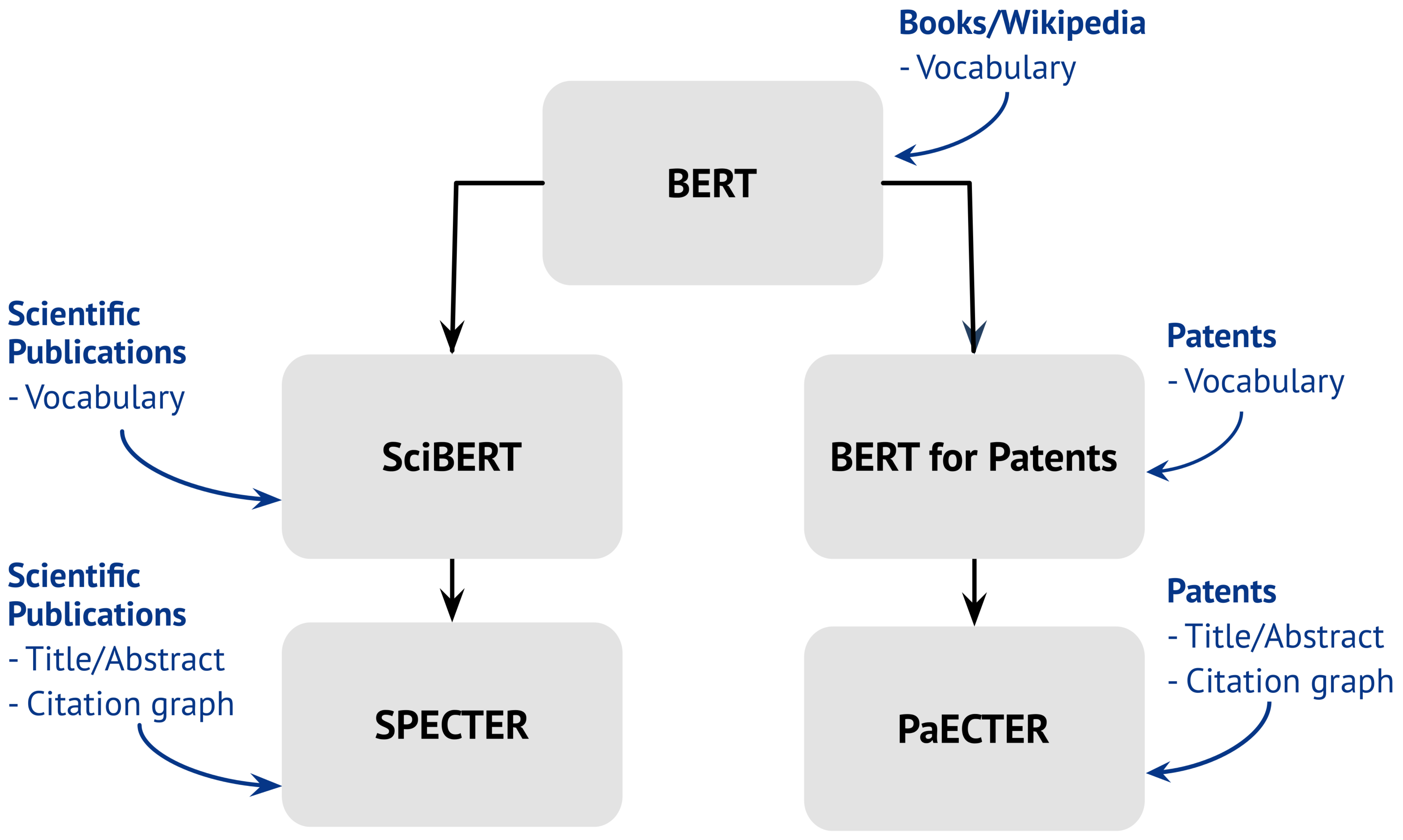}
\end{figure}

We strive to fill a gap in the literature by providing a model that combines domain-specific vocabulary, citation-induced similarity training, and public availability. To this end, we adopt the contrastive learning technique akin to the SPECTER/SPECTER2 models. Our model is built on ``BERT for Patents'', with the training objective to distinguish cited from non-cited patents based on semantic similarity to a focal patent. Fig.~\ref{fig:bert-and-specters} shows how the different models relate to each other.

In total, we use 300,000 patent families with a family member in the EPO, and generate five triplets for each.

In comparisons with state-of-the-art models available at the time of training, PaECTER shows superior capabilities in predicting semantic similarity. PaECTER supports downstream tasks in the patent domain, such as topic classification, citation prediction, and recommendation. As such, PaECTER is particularly helpful in the time-consuming patent examination task.

Since PaECTER's release, several studies have benchmarked against PaECTER on a variety of tasks. We report on the benchmarking exercises of \cite{ganguli_patent_2024} and \cite{ayaou_patenteb_2025} in section ``External Evaluation.''

There is also a growing body of work that provides alternative models for patent ranking and retrieval. We summarize these developments to situate our contribution within the current landscape rather than to claim priority over later efforts. Recent model developments include the patembed model family by \cite{ayaou_patenteb_2025}, the unbranded model by \cite{daniell_efficient_2025} as well as the PAI-NET by \cite{lee_pai-net_2025}. PAI-NET \cite{lee_pai-net_2025} also uses a contrastive learning approach, but trains on examiner-added patent citations in patents granted by the USPTO or the KIPO. The patembed family \cite{ayaou_patenteb_2025} is also built on BERT for Patents but is trained jointly across multiple retrieval and classification tasks and extends the input context window. The model of \cite{daniell_efficient_2025} departs from text encoders altogether, as they represent each patent document as a graph of features and their relationships that a Graph Transformer embeds for dense retrieval.

\section*{Training data}

Our training data involves 300,000 English-language patent applications filed with the European Patent Office (EPO) from 1985 to 2022.

We specifically select EPO patents from these families for two critical reasons. First, references cited in EPO search reports are under the full control of examiners, mitigating any strategic bias typically associated with citations made by inventors or patent attorneys. Second, these references are classified with a unique emphasis on their contribution to the claimed invention \cite{webb_analysing_2005, wada_when_2020}, a feature not present in the United States Patent and Trademark Office (USPTO).

\subsection*{Objective}
The primary goal of our model is to learn textual similarity and dissimilarity. By feeding the model examples of similar and dissimilar documents in relation to a focal document, the model autonomously learns optimal weights that minimize the loss function. Triplet margin loss is an effective approach in this context, where the optimizations of the weights are guided by maximizing the differences between similar and dissimilar examples. 

Our training dataset comprises triplets of patents with a focal patent, a similar patent (a positive example), and a dissimilar patent (a negative example). Similar patents are those that the focal patent cites directly with a specific citation category. Dissimilar patents are those it does not cite, subject to further refinements.

The triplet margin loss function is defined as:
\begin{equation}
    \max\{(\lVert V_F - V_P \rVert_2 - \lVert V_F - V_N \rVert_2 + m), 0\}
    \label{eq:loss_fnc}
\end{equation}
where $\lVert.\rVert_2$ represents the $L_2$ norm distance, and $V_F, V_P, V_N$ are numerical representations for the focal patent ($P_F$), positive patent ($P_P$), and negative patent ($P_N$), respectively. The margin variable ($m$) ensures that the positive patent is closer to the focal patent than the negative one by at least a specified margin.

\subsection*{Focal patents} We utilize PATSTAT 2023 Spring version, a proprietary database administered by the European Patent Office (EPO), containing bibliographical and legal event data for patents worldwide. We selected 300,000 patent applications as focal patents following specific criteria:
\begin{enumerate}
    \item They are either EP-direct or Euro-PCT applications.
    \item They were filed between 1985 (We exclude patents filed prior to 1985 due to the ongoing transition to the EPO during that period) and 2022.
    \item They have been assigned a Cooperative Patent Classification (CPC).
    \item They have a minimum of two backward citations of category \textit{X/Y/I}, or at least one backward citation of category \textit{X/Y/I} and one of category \textit{A}.
    \item Their referenced patents collectively cite at least two other patents.
\end{enumerate}

Given our focus on English language training, all patents, including focal, positive, and negative ones, must have an English title and an English abstract. In cases where this condition is not met, we strive to substitute it with the most fitting member of the same DocDB family. A DocDB family refers to those patent applications and grants that originate from the same set of priority filings but seek protection in different jurisdictions. We utilize a patent document from another authority according to the following priority rule:
\begin{equation*}
WO > US > GB > CA > AU > DE >
CN > TW > KR > FR > JP
\end{equation*}

In total, 1,358,264 patent documents satisfy these criteria. From those, we randomly sample 300,000 documents after deduplicating based on text (title concatenated with abstract). Since all our focal patents are granted EP patents, multiple patents belonging to the same DocDB family may arise only due to divisional patent filings \cite{harhoff_patent_2016}. No DocDB family exists more than once in our dataset though.

Finally, we split the training dataset into 85:15 ratio, such that all triplets for the same focal document remain together. The 15\% fraction is reserved for validation purposes during the training process.

\subsection*{Positive citations}

Positive citations are sampled with replacement from the eligible pool for each focal patent, based on specific citation categories (Table \ref{tab:citation_category}). Of these categories, \textit{X}, \textit{Y}, \textit{I}, and \textit{A} are particularly relevant, as they indicate a strong similarity or relevance of the corresponding cited patent to the focal patent.

\begin{table}[ht]
  \centering
  \caption{Citation Categories Provided by the EPO}
  \label{tab:citation_category}
  \begin{tabular}{cp{5.5cm}}
\toprule
Category & \multicolumn{1}{c}{Meaning} \\
\midrule
\textbf{X} & Particularly relevant if taken alone - prejudicing novelty. \\
\textbf{I} & Particularly relevant if taken alone - prejudicing inventive step. \\
\textbf{Y} & Particularly relevant if combined with another document of the same category. \\
\textbf{A} & Documents defining the state of the art and not prejudicing novelty or inventive step. \\
O & Non-written disclosure. \\
P & Intermediate document. \\
T & Theory or principle underlying the invention. \\
E & Earlier patent application, but published after the filing date of the application searched (potentially conflicting patent documents).  \\
D & Document cited in the application. \\
L & Document cited for other reasons. \\
\& & Member of the same patent family, corresponding document \\
\bottomrule
\end{tabular}
  \par
  \vspace{0.2cm}
  \raggedright
  \textbf{Source} From \cite{webb_analysing_2005}. Categories in bold font are the ones we are using. Category ``I'' has been introduced in April 2011 (see \href{https://register.epo.org/help?topic=citations&lng=en}{https://register.epo.org/help?topic=citations\&lng=en}). 
\end{table}

Categories \textit{X}, \textit{Y}, and \textit{I} reflect citations where the cited patent is considered highly similar and potentially prejudicial to the novelty or inventive step of the focal patent \cite{wada_when_2020}. Although not novelty-destroying, citations of kind \textit{A} reflect the state of the art for the focal patents. Accordingly, we treat cited patents of kinds \textit{X}, \textit{Y}, \textit{I}, and \textit{A} as similar to the focal patents.

The sampling with replacement introduces duplicates at the level of the focal patent. Duplicates may also enter if they are cited by two included patents. The positive samples of the complete training set have 901,587 unique patent application IDs out of 1,500,000 rows. However, $60\%$ of positive patents are used only once.

\subsection*{Negative citations}
Negative patents, representing dissimilarity with the focal patents, are selected with a nuanced approach. First, they must come from the universe of all patents ever cited by any EPO patent, which was filed during the 1985--2022 period. Second, at least the abstract is available in English through the cited patent itself or its best substitute. Third, a CPC category must exist.

We then group the negatives into easy negatives and hard negatives:
\begin{itemize}
    \item \textbf{Easy Negatives:} Non-cited patents in the same CPC class and published within five years prior to the focal patent, and not indirectly cited.
    \item \textbf{Hard Negatives:} Patents cited by the focal patent's backward citations but not by the focal patent itself.
\end{itemize}

\begin{figure}
  \centering
  \caption{Positive, easy and hard negatives in patent training selection}\label{fig:pos_neg_citations}
  \includegraphics[width=7cm]{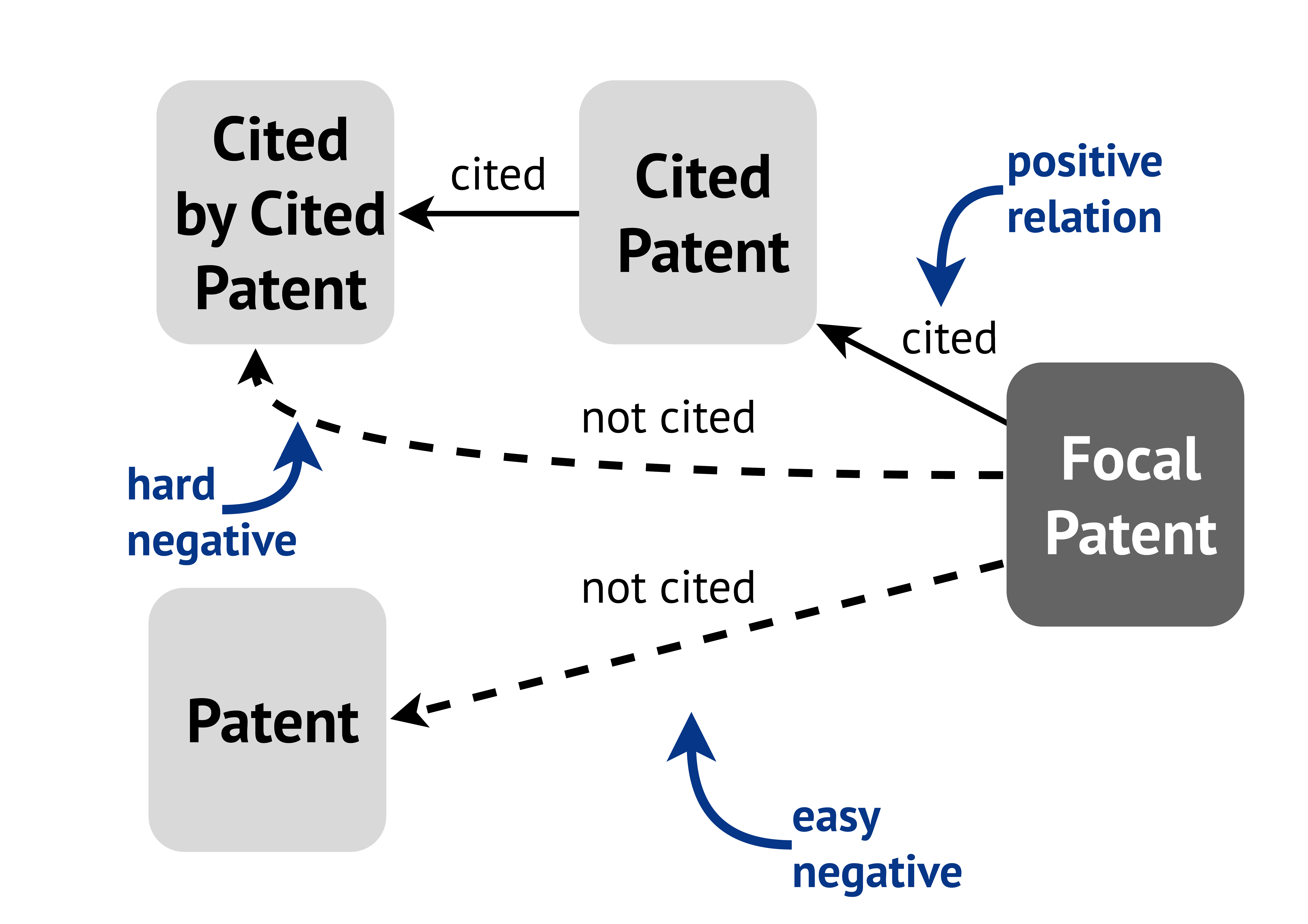}
\end{figure}

Fig.~\ref{fig:pos_neg_citations} illustrates the selection process. The number of all potential easy negatives is 6,484,216; there was no focal patent that did not meet the requirements of at least three easy negatives.

With the selection of the negatives, there are also some duplicates. The negative samples of the complete training set have 1,251,737 unique patent application IDs out of a total of 1,500,000. In retrospect, we discovered that 30 hard negatives were positive citations of patent siblings. These patent siblings were divisional filings, which are rare at the EPO \cite{harhoff_patent_2016}.

\subsection*{Test dataset}
For the test dataset, we construct 1000 samples consisting of 1 focal patent with 5 positive citations and 25 negative citations. The negative citations are a random sample of 10 hard negatives and 15 easy negatives of a patent. Since the dataset is smaller, only focal patents were selected where there were enough positive and negative examples. We do not sample with replacement. 

The test dataset has no overlap with the training data. None of the focal, negative, or positive patent selections are present in the training dataset.

The test sample is furthermore intentionally representative. 976 of our test documents belong to a multi-authority patent family. 921 of these include a US sibling.

\section*{Training}

We build our model utilizing the sentence transformers architecture \cite{reimers_sentence-bert_2019} and HuggingFace framework \cite{wolf_transformers_2020}. At its core, the model uses BERT for Patents \cite{srebrovic_leveraging_2020}, which is connected to an input module and a subsequent document-level embedding module.

The input module concatenates a patent's title and abstract and passes the concatenated text to the BERT for Patents model. Within our text corpus, 1,753 abstracts contain [UNK] tokens using BERT for Patents. For comparison, the number of abstracts with [UNK] tokens using BERT is 1,743.

Because of the architecture of BERT models, only the first 512 tokens (about 393 words) are used. In our corpus, the mean number of tokens per document is 159 with 4,895 documents (0.25\%) being above the 512 token limit.

All the 1,024-dimensional tokens of the output layer of the BERT for Patents model subsequently go to the document-level embedding module, which produces the mean embedding of all these tokens as the final numerical representation for the given input patent text. We experimented with the mean embedding of the output tokens from the BERT for Patents model; it outperformed the embedding of the [CLS] token in the similarity learning task. Hence, we do not use the [CLS] token as the final representation of any given text.

During training, we feed each triplet to the model. It then optimizes the triplet loss function (Eq. \eqref{eq:loss_fnc}) by comparing the numerical representations of all three documents in that triplet. 

We do minimal hyperparameter-tuning for training, following the leads of SPECTER and SEARCHFORMER. 
Post-tuning, we use a learning rate of $1e$-$5$, 10\% of the training data for warmup steps, and a triplet loss margin ($m$) of 1 for training, in addition to the \textit{Decoupled Weight Decay Regularization} optimizer \cite{loshchilov_decoupled_2019}.

We train the model for four epochs, though performance plateaus after the second. 
Each training epoch spans approximately 20 hours, with validation occurring every 4000 steps.

We use a batch size of 4 (the maximum supported by the GPU memory) and gradient accumulation of 8 on four NVIDIA A100-SXM4-40GB GPUs for faster training and achieving an effective batch size of 128. 
To train the model with multiple GPUs, we have adapted the code base of the sentence transformers package slightly, and also utilized HuggingFace's Accelerate library.

\section*{Evaluation}\label{sec:Evaluation}






\subsection*{Performance evaluation}

We conduct a rank-aware evaluation of our trained model, PaECTER, and compare it against several other models and algorithms on our test dataset. 
The baselines of our evaluation include the BM25 algorithm \cite{robertson_simple_1994,robertson_probabilistic_2009}, BERT, SciBERT, SPECTER, SPECTER version 2 (BASE), BERT for Patents, PatentSBERTa \cite{bekamiri_patentsberta_2024} and Patembed \cite{ayaou_patenteb_2025}. The latter comes in several flavors which users steer via prompts: we use the same retrieval (IN), the different retrieval (OUT) and the mixed retrieval (MIXED). We also add more recent general-purpose text embedding models such as GTE \cite{li_towards_2023}, BGE \cite{xiao_c-pack_2024}, and E5 \cite{wang_text_2024}. 

We use four evaluation metrics. The first, \textit{Rank First Relevant} (RFR), is the position of the first actual reference patent within the ranked list of related patents for each sample. Secondly, we use \textit{Mean Reciprocal Rank @ 10} (MRR@10), which averages the reciprocal ranks of the first correctly predicted patent among the top 10 ranked patents, averaged over all test samples. Thirdly, we use \textit{Normalized Discounted Cumulative Gain @ 10} (nDCG@10), which, treating each candidate as simply relevant or non-relevant, measures the discounted cumulative gain of the top 10 ranked patents normalized against that of an ideal ranking, averaged across all test samples. Finally, we use \textit{Mean Average Precision} (MAP), which measures the average precision of identifying all five actual reference patents within the ranked list of 30 candidate patents for each sample, and then averages these precision scores across all samples.

Moreover, since pooling strategies can potentially affect model performance, we consider two widely used pooling strategies during evaluation, namely CLS and mean. 
Table \ref{table:rank_evaluation_patent} presents the performances of all models under consideration for both strategies.

In Table \ref{table:rank_evaluation_patent}, all four metrics point to the superiority of the PaECTER model over all other models, regardless of the pooling strategy used. In particular, our trained model with mean pooling achieves higher accuracy over BERT for Patents by $7.85$ points (significant, $p<.001$) in MAP and $7.76$ points (significant, $p<.001$) in MRR@10.

\begin{figure}
    \centering
    \caption{ECDF for the distribution of RFR scores across different models\label{fig:rfr}}
    \includegraphics[width=0.9\textwidth]{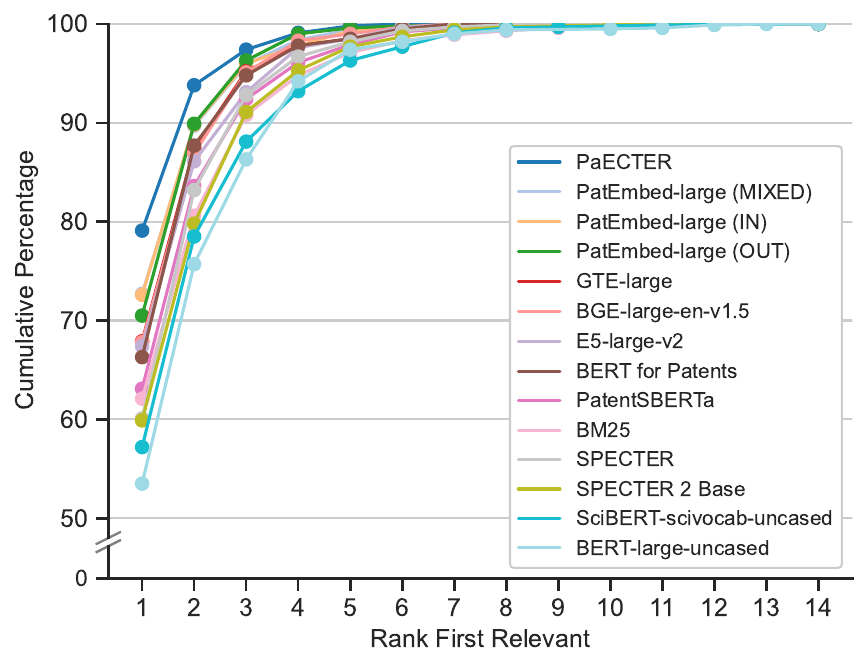}
    \par
    \raggedright
    \textbf{Note:} The distribution of \textit{Rank First Relevant} (RFR) scores for PaECTER and other models used in comparison. All models use their best pooling strategy, as determined from Table \ref{table:rank_evaluation_patent}.
\end{figure}

Furthermore, Fig.~\ref{fig:rfr} illustrates that, in terms of the RFR metric, our model consistently ranks the first relevant patent among the top positions compared with other models.

\begin{table}
  \centering
  \caption{Rank-aware evaluation of several models}
  \label{table:rank_evaluation_patent}
  \begin{tabular}{l cc cc cc cc}
    \toprule
    & \multicolumn{2}{c}{Avg. RFR} & \multicolumn{2}{c}{MRR@10} & \multicolumn{2}{c}{nDCG@10} & \multicolumn{2}{c}{MAP} \\
    \cmidrule(lr){2-3} \cmidrule(lr){4-5} \cmidrule(lr){6-7} \cmidrule(lr){8-9} Model & CLS & Mean & CLS & Mean & CLS & Mean & CLS & Mean \\
    \midrule
    BERT-large-uncased & 2.92 & 1.98 & 58.95 & 71.06 & 43.92 & 60.94 & 38.17 & 51.69 \\
    SciBERT-scivocab-uncased & 3.07 & 1.91 & 58.80 & 73.45 & 43.30 & 60.35 & 37.76 & 51.31 \\
    PatentSBERTa & 1.68 & 1.71 & 77.89 & 77.26 & 67.01 & 67.89 & 56.97 & 57.56 \\
    SPECTER & 1.74 & 1.70 & 77.04 & 76.40 & 65.52 & 65.98 & 55.86 & 56.10 \\
    SPECTER 2 Base & 1.79 & 1.68 & 75.47 & 77.96 & 65.65 & 67.08 & 56.12 & 56.96 \\
    E5-large-v2 & 1.61 & 1.58 & 80.26 & 80.63 & 69.06 & 70.08 & 58.94 & 59.78 \\
    BERT for Patents & 2.44 & 1.55 & 67.14 & 80.49 & 51.72 & 70.52 & 44.44 & 60.32 \\
    GTE-large & 2.25 & 1.54 & 71.12 & 81.27 & 54.13 & 70.81 & 46.64 & 60.25 \\
    BGE-large-en-v1.5 & 1.56 & 1.53 & 80.37 & 81.18 & 69.82 & 70.83 & 59.57 & 60.51 \\
    patembed-large (IN) & 1.53 & 1.45 & 81.22 & 84.12 & 70.25 & 74.67 & 60.13 & 63.80 \\
    patembed-large (OUT) & 1.97 & 1.45 & 70.26 & 83.19 & 61.78 & 74.60 & 51.94 & 63.94 \\
    patembed-large (MIXED) & 1.55 & 1.44 & 81.50 & 84.19 & 69.64 & 74.87 & 59.80 & 63.91 \\
    PaECTER & \textbf{1.32} & \textbf{1.31} & \textbf{88.03} & \textbf{88.25} & \textbf{77.39} & \textbf{78.68} & \textbf{66.91} & \textbf{68.17} \\
    \bottomrule
\end{tabular}
  \par
  \vspace{1pt}
  \raggedright
  \textbf{Note:} We show the performance of the PaECTER model in terms of \textit{Rank First Relevant} (RFR),  \textit{Mean Reciprocal Rank} (MRR) at $k=10$, normalized Discounted Cumulative Gain at $k = 10$ (nDCG) as well as \textit{Mean Average Precision} (MAP), and compare it with other models. The MAP metric uses 5 positive and 25 negative patents (10 hard negatives and 15 easy negatives). RFR and MRR@10 use only the first positive. Scores are rounded to two decimal places. Bold values represent the optimum in each column.
\end{table}

\subsection*{Comparison with SEARCHFORMER}

SEARCHFORMER is another domain-specific model designed by \cite{vowinckel_searchformer_2023} for patent examination. It was trained specifically to assist EPO patent examiners in prior art search. Unlike PaECTER, it was trained on claims texts rather than titles and abstracts. SEARCHFORMER is a closed-source language model and the training dataset is confidential.

To assess the relative performance of PaECTER, we compare its rank predictions with those of SEARCHFORMER using SEARCHFORMER’s test dataset. The goal of this evaluation is to suggest relevant prior art given the first claim of a patent.

From the authors of SEARCHFORMER, we obtained about 90\% of the original SEARCHFORMER test dataset, excluding pending applications (which are confidential). This subset includes 1,795 focal patents and a total of 1,154,557 pairs. Each focal patent is associated with 1--14 positive citations and between 197 and 955 negative citations. 

\begin{table}[ht]
    \centering
    \caption{Rank-aware evaluation of PaECTER against SEARCHFORMER using the SEARCHFORMER dataset\label{tab:rank_evaluation_searchformer}}
    \begin{tabular}{lccc}
    \toprule
    Model & Avg. RFR & MAP & MRR@10 \\
    \midrule
    SEARCHFORMER & 58.73 & 9.38 & 15.33 \\
    \addlinespace
    PaECTER & \textbf{49.66} & \textbf{11.13} & \textbf{17.93} \\
    \bottomrule
\end{tabular}
    \par
    \vspace{1pt}
    \raggedright
   \textbf{Note:} We show the performance of the SEARCHFORMER and PaECTER models in terms of \textit{Rank First Relevant} (RFR), \textit{Mean Average Precision} (MAP), and \textit{Mean Reciprocal Rank} (MRR) at $k=10$. The MAP metric uses 5 positive and 25 negative patents (10 hard negatives and 15 easy negatives). RFR and MRR@10 use only the first positive. Scores are rounded to two decimal places. Bold values represent the optimum in each column.
\end{table}

Table \ref{tab:rank_evaluation_searchformer} portrays the results of this exercise. It shows that PaECTER outperforms SEARCHFORMER in all metrics. Based on the RFR metric, we see that on average, an examiner would have to inspect nearly 50 documents suggested by PaECTER to find relevant prior art, which is 9 documents fewer than with SEARCHFORMER.

That PaECTER outperforms SEARCHFORMER may reflect different training strategies. SEARCHFORMER uses claims as a proxy for citation prediction, whereas PaECTER uses titles and abstracts but for the same purpose. While patent claims define the legal scope of protection and are central to novelty assessments, our evaluation task differs from a full patentability assessment. Both our benchmark and the SEARCHFORMER benchmark evaluate citation prediction through textual similarity, rather than detailed claim interpretation. PaECTER’s stronger performance relative to SEARCHFORMER suggests that titles and abstracts contain sufficient information to capture many of the semantic signals underlying examiner citation decisions, making them effective for identifying technologically related prior art. More specifically, our evaluation setup rewards the retrieval of relevant and technologically related patents rather than the precise matching of technical features embodied in patent claims. PaECTER appears particularly well suited for the former, where technological relevance is the primary objective.

At the same time, this finding should not be interpreted as evidence that claims are unnecessary for patent examination or prior-art analysis. Applications requiring legal scrutiny, novelty assessments, inventive-step evaluations, or Freedom-to-Operate analyses are likely to benefit from incorporating claims and potentially the full patent specification in a more extensive exercise.

\subsection*{External evaluation\label{sec:ext_eval}}
\cite{ganguli_patent_2024} have tested the PaECTER, among other models, on several tasks involving independent claims of US patents published between 1850 and 2023. Other models include S-BERT and GTE, next to a TF-IDF representation as a baseline model. \cite{ganguli_patent_2024} compare these models in three tasks: (i) an interference task, (ii) a human annotation task, and (iii) a technology classification task.

The interference task was to connect two (pending) patents on the same invention by two independent inventors. The task used decisions made by patent examiners between 1998 and 2014. Results show that PaECTER outperforms all other models, though GTE follows second by a small margin.

In the human annotation task, the models had to replicate non-expert human judgment on which patent was more similar in a pair-wise comparison. The task included patents published during the 1880--1920 period. In this task, PaECTER comes third with GTE being best. As the authors note, this is likely due to the historical data involved in the test.

Finally, the technology classification task involved predicting whether a pair of patents belongs to the same CPC section (9 in total) or CPC class (123 in total). The test included patents published from the 1850--2023 period. Depending on the metric (ROC AUC or PR AUC), PaECTER ranks first or second (behind S-BERT) in predicting whether two patents belong to the same CPC class. It ranks second or third in predicting shared CPC section membership, though all models perform similarly on this task.

\cite{ayaou_patenteb_2025} have also tested PaECTER, among other models, as a baseline in their PatenTEB benchmark, a suite of 15 patent-specific tasks constructed from roughly 2.06 million examples drawn from multi-jurisdictional patent families. The benchmark groups the tasks into four families: retrieval, classification, paraphrase detection, and clustering. Other models include the authors' own patembed family, general-purpose encoders such as GTE-ModernBERT and Qwen3-Embedding, and patent-specialized baselines such as PatentSBERTa and BERT for Patents.

The retrieval family comprises eight tasks: three symmetric document-to-document tasks (same-domain, mixed-domain, and cross-domain) in which relevance is defined by citation links, and five asymmetric fragment-to-document tasks (title-to-full, problem-to-full, problem-to-solution, effect-to-full, and effect-to-substance) that query full documents using only a fragment. Performance is measured by nDCG@10, a rank-aware metric. The three symmetric tasks are the ones closest to PaECTER's own training objective of citation-supervised prior-art retrieval, and on these PaECTER is the strongest model after the purpose-built patembed family, ahead of both general-purpose encoders (e.g., 0.421 vs. 0.395 for Qwen3 and 0.363 for GTE-ModernBERT on the same-domain task). On the asymmetric tasks, which lie outside its design, PaECTER trails the general encoders, most clearly on title-to-full (0.454 against 0.689--0.738).

The classification family covers three tasks: assigning patents to one of 109 IPC3 technology classes, predicting a patent's citation-timing trajectory, and predicting the direction of a citation between two patents. Performance is measured by the macro-F1. Here PaECTER averages 0.536, resulting in a third place. This reflects that its document-level embeddings capture technological domain information well even though the model was not trained for classification.

The paraphrase family consists of two tasks that test whether the model recognizes problem or solution segments as semantically equivalent, with positive pairs sampled from same-domain citation pairs. Performance is measured by the Pearson correlation coefficient. PaECTER reaches 0.753 on these tasks, below the patembed family and somewhat above the general encoders.

The clustering family comprises two evaluation-only tasks that group patent families by their full set of IPC codes and by inventor identity. Performance is measured by V-measure. PaECTER scores 0.621, among the best of all evaluated models and slightly above patembed-large. This result is somewhat surprising given that clustering is a task it was never trained for.

\subsection*{Ablation method}
We perform an ablation study whereby we include only title and CPC classes, but not abstracts, in the training of the model. The Cooperative Patent Classification (CPC) is the most used technology classification system and available for almost every patent. It consists of sections, classes, subclasses, main groups and subgroups, thereby providing a hierarchical or nested classification. At the lowest granularity, there are well more than 250k CPC subgroups. Most patents belong to several CPC subgroups (\cite{lobo_sources_2019} found an average of 5 for patents granted by the USPTO between 1975 and 2018).

Since not every patent has CPC codes, the training, validation and test sets decrease somewhat in size as compared to the original application. For example, about 103k positive samples from the original training set lack CPC codes.

\begin{table}
    \centering
    \caption{Ablation method with CPC codes instead of abstracts\label{tab:ablation}}
    \begin{tabular}{lcccccc}
    \toprule
    \multirow{2}{*}{Model/Test} & 
    \multicolumn{2}{c}{Avg. RFR} & 
    \multicolumn{2}{c}{MAP} & 
    \multicolumn{2}{c}{MRR@10} \\
    \cmidrule(lr){2-3} \cmidrule(lr){4-5} \cmidrule(lr){6-7}
    & CLS & Mean & CLS & Mean & CLS & Mean \\
    \midrule
    Original Model / Original Testset & \textbf{1.32} & \textbf{1.31} & \textbf{66.91} & \textbf{68.17} & \textbf{88.03} & \textbf{88.25} \\
Original Model / Ablated Testset & 2.01 & 1.92 & 56.59 & 57.63 & 71.87 & 72.1 \\
Ablated Model / Original Testset & 1.59 & 1.48 & 57.37 & 63.59 & 79.23 & 82.47 \\
Ablated Model / Ablated Testset & 1.88 & 1.83 & 59.4 & 60.55 & 73.96 & 75.04 \\
\bottomrule
\end{tabular}

    \par
    \vspace{1pt}
    \raggedright
    \textbf{Note:} We show the performance of the PaECTER model and the ablated PaECTER model in terms of \textit{Rank First Relevant} (RFR), \textit{Mean Average Precision} (MAP), and \textit{Mean Reciprocal Rank} (MRR) at $k=10$. The MAP metric uses 5 positive and 25 negative patents (10 hard negatives and 15 easy negatives). RFR and MRR@10 use only the first positive.  Scores are rounded to two decimal places. Bold values represent the optimum in each column. The ablated test set has about 5k samples less than the original test set.
\end{table}

Table \ref{tab:ablation} presents the results. We perform the test separately for training data with CPC codes instead of abstracts, as well as for the ablated test set. For all the metrics and pooling methods we find that the original model (trained on abstracts) outperforms the ablated model.

\section*{Conclusion}

The PaECTER model provides a valuable resource for researchers and practitioners analyzing patent text. It is freely available on HuggingFace and fast to deploy.

PaECTER's main task is that of a citation recommender, i.e., it excels at retrieving related patent literature for instance during the examination process. Though PaECTER can in principle be used for a variety of tasks (semantic distance, classification/clustering, etc.), we recommend users to benchmark several models on these tasks.

\section*{Limitations}

The PaECTER model has several limitations that are important to consider:

\paragraph{Token length constraints.} The underlying BERT architecture processes sequences of up to 512 tokens. As a result, PaECTER was trained on patent titles and abstracts, rather than full descriptions or claims, which are central to defining a patent’s legal scope. While the model captures general semantic aspects of patents, it may be less effective at representing their fine-grained technical details.

\paragraph{Language scope.} The model was trained exclusively on English-language patent texts. It is therefore not directly applicable to non-English patents. Users working with non-English texts should apply high-quality machine translation prior to inference.

\paragraph{Temporal relevance.} Because the training data reflects existing and past technologies, the model may underperform on emerging innovations not yet well-represented in the dataset. In particular, it may struggle to identify or classify technologies requiring new Cooperative Patent Classification (CPC) categories in the future. This is left for future analysis.

\section*{Acknowledgements}
We thank the team at the Max Planck Computing and Data Facility (MPCDF) for allowing us to use their high-performance computing cluster. Finally we thank participants of the European Patent Office's ARP Workshop 2023 (in particular our discussant, Gaétan de Rassenfosse) as well as the 2024 Munich Summer Institute for helpful comments.

\section*{Funding}
Our work has been enabled by grant 2023/8401 from the European Patent Office under their Academic Research Programme (EPO-ARP) 2021. The funders had no role in study design, data collection and analysis, decision to publish, or preparation of the manuscript.

\section*{Data availability}
The model is available at \url{https://huggingface.co/mpi-inno-comp/paecter}. The training dataset is available at \url{https://huggingface.co/datasets/mpi-inno-comp/paecter_dataset}. The validation data set and task is available at \url{https://github.com/Max-planck-innovation-competition/patents-to-patents}.

\section*{Competing interests}
The authors have declared that no competing interests exist.

\section*{Author contributions}
Conceptualization: M.G.; M.R.; S.E.; D.H. Data curation: M.R. Funding acquisition: M.R.; D.H. Investigation: M.G. Methodology: M.G.; M.R.; S.E. Project administration: E.B.; D.H. Resources: M.G.; S.E. Software: M.G. Supervision: M.R. Validation: S.E.; E.B. Visualization: E.B. Writing -- original draft: M.G.; M.R.; S.E.; E.B.; D.H. Writing -- review \& editing: M.R.

\bibliography{references}

\end{document}